# Design and Implementation of a Cloud Computing Security Assessment Model Based on Hierarchical Analysis and Fuzzy Comprehensive Evaluation

Yihong Zou

Intent Driven Network, Amazon Data Services, Inc, Cupertino, California, 95014, US
zyh.gc.2024@gmail.com

**Abstract.** At the rapid pace of technological evolution, the emerging cloud computing technology has promoted the digitalization and business innovation of the enterprise in all industries due to its advantages of data storage and service mode. Nevertheless, given the swift progress in cloud computing services, the security problems have gradually appeared. The data breach and cyber attack happen frequently, which cause huge losses to enterprises and individuals. These issues have gradually become important constraints for the popularization of cloud computingTo tackle the problems outlined above, this paper constructs a security evaluation framework for cloud computing services, integrating the Analytic Hierarchy Process (AHP) with the Fuzzy Comprehensive Evaluation method. By applying this scientific and systematic methodology, the framework enables enterprises and individuals to better apprehend the security posture of cloud services, thereby fostering the healthy evolution of the entire industry.

**Keywords:** Cloud Computing Security Assessment, Hierarchical Analysis Method, Fuzzy Integrated Assessment Approach

## 1. Introduction

Emerging from the tide of informatization and digitization, as an innovative information technology, cloud computing has quickly developed into an important driving force that promotes the digitalization of other businesses due to its characteristics of high efficiency, strong mobility and scalability. Now, the cloud computing technology can effectively reduce enterprises' IT cost and improve the speed of business innovation and response to the market by providing them with needed computational resources and services on demand. With the deep and widespread application of cloud computing, security problems have gradually appeared and have become the key bottleneck that restricts the further development of cloud computing. The security problems of cloud computing services are multi-level and include data breach, network attack, identity authentication, etc. They not only have a huge impact on the normal operation of businesses, but also threaten users' privacy and property safety. In response to the demand for evaluating the security of cloud computing, scholars and institutions at home and abroad have conducted in-depth investigations and implementations. However, most of the existing assessment methods have problems such as imperfect evaluation indicators and inaccurate assessment results, which cannot meet the demand for further improving the security level of cloud computing services. This research studies a cloud computing security evaluation model that constructs an complete cloud computing security evaluation indicator system by combining hierarchical analysis approach and fuzzy overall assessment technique to improve the accuracy and reliability of the assessment results and provide a scientific basis for the decision-making of cloud computing service providers and users[1]. Furthermore, the proposed framework aligns with national regulatory requirements and standards for cloud computing security, particularly those pertaining to security capabilities and the classified protection assessment. This alignment offers robust technical support for the continued advancement of the domestic cloud computing industry.

## 2. Related Research

### 2.1 Hierarchical Analysis Method

The Hierarchical Analysis Method (HAM), introduced by Professor T.L. Saaty in 1970, is an effective structured decision-making analysis tool that breaks down intricate decision-making challenges into numerous smaller issues or elements through the construction of a hierarchical model[2]. It determines the relative importance of each element through pairwise comparisons and ultimately offers a quantitative foundation for making decisions. The method includes the following steps: a. Constructing a Hierarchical Framework Model: Based on their interconnections, the decision goals, evaluation criteria, and decision alternatives are categorized into the top tier, intermediate tier, and bottom tier, and a hierarchical framework diagram is depicted. b. Building Evaluation Matrices: For consecutive levels, the superior level is termed the objective level, while the inferior level is referred to as the criterion level. By conducting comparative assessments, the relative significance of the criterion level to the objective level is established, resulting in the creation of an evaluation matrix. c. Weight Determination: Techniques like the eigenmethod or the reciprocal scaling approach are employed to compute the eigenvector of the evaluation matrix, which is subsequently standardized to derive the weights of each criterion. Following a sequence of computations, a consistency validation of the evaluation matrix is necessary to guarantee the rationality and credibility of the weights[3]. Additionally, the weight values of all factors at a certain level relative tin relation to the overarching objective are calculated. Tummala R. believes that applying AHP in the field of cloud computing security assessment has many advantages. It can systematically handle multi-criteria, multi-level evaluation problems and provide a comprehensive and hierarchical evaluation framework for the security of cloud computing services.

### 2.2 Fuzzy integrated assessment approach

The fuzzy comprehensive evaluation method provides a valuable methodology for addressing problems characterized by fuzziness and uncertainty. By establishing the fuzzy set and membership function, it quantifies evaluation indicators, the evaluation result is more accurate and reliable. YANG Z H. believes that the application of fuzzy integrated assessment approach should clarify the purpose of evaluation and objects of evaluation as well as the range of indicators of evaluation, establish the fuzzy relation matrix, determines the weights of indicators based on subjective weighting method or objective weighting method or subjective and objective weighting method, calculate membership degree, then do fuzzy operation to get weighted score matrix, at last, do normalization to get the final evaluation result[4]. In the security assessment of cloud computing, the fuzzy integrated assessment approach can solve the shortcoming that the traditional evaluation approach can't handle fuzzy information, reflecting the security situation of cloud computing service more accurately. However, the accurate definition of membership functions for fuzzy sets and the rational assignment of indicator weights continue to present significant challenges[5]. Besides, the calculation process of fuzzy integrated assessment approach is rather complicated, only by using the professional software or algorithm, the method can be implemented, so it's not convenient to popularize in practice.

**2.3 Cloud Computing**

As a new pioneering technology in information technology field, cloud computing features outstanding advantages in gaining scale effects and extending resource space. In terms of international situation, cloud computing not only promotes the research of IaaS platform, such as Google App Engine and Microsoft Azure, but also gives birth to SaaS products, such as Salesforce SalesCloud. With the commercially available service platform, such as Alibaba Cloud, Sina Cloud, etc., enterprisedealing methods and individuals' digital consumption mode are changing rapidly. The user data and business logic running on these cloud computing platforms are consuming a lot of computational resources. Any security hole or attack would cause huge losses, so the security issues of cloud computing platforms are always a concern. With the popularization and promotion of cloud computing technology, the research of cloud computing security assessment has made rapid development. The method of assessment can be roughly classified into qualitative assessment method, quantitative assessment method and mixed assessment method. These methods have advantages and disadvantages in different situations. Based on Service Level Agreement (SLA), Jiang et al. used Goal-Question-Metric (GQM) method to establish indicator set and scored cloud service provider's security ability by measuring the gap with "best", "worst" and "required" levels [6]. In the research of Li et al., indicators were directly used from Cloud Control Matrix (CCM), and the Technique for Order Preference by Similarity to Ideal Solution (TOPSIS) method was used to evaluate the security performance of candidate cloud service provider [7]. Liu, on the other hand, combined two types of information: objectively, the monitored values of QoS attributes, both integrated through the TOPSIS method to derive security capability scores for cloud service providers.

# 3. Method

**3.1 Indicator Determination**

The selection of evaluation indicators is the basis for establishing an evaluation system and the key to the accuracy and completeness of the evaluation results. The cloud computing security evaluation indicators in this paper were constructed based on a set of key principles—scientific and systematic, hierarchical and independent, comparable and operable—with the objective of comprehensively capturing the platform's security posture. a. Scientificity and Systematicness. The selection of evaluation indicators should be based on the deep research of cloud computing security, fully consider all kinds of factors that affect the security of the platform, so that the results of the evaluation can reflect the security status of the cloud computing platform. b. Hierarchy and Independence: Evaluation indicators should be synthesized and decomposed according to different attributes to form an indicator system with different levels. Indicators at each level should be mutually independent, avoiding overlapping and inclusion relationships to ensure the accuracy and reliability of the evaluation results. c. Comparability and Operability: Evaluation indicators should be comparable and operable, facilitating horizontal comparisons of security status among different cloud service providers. At the same time, indicators should be easy to measure and collect data to facilitate the implementation of the evaluation work[8]. When using the Delphi method to measure the risk level of each risk source, to avoid different experts giving different risk level assessment results due to differences in knowledge and experience, we established a five-category risk level classification based on Neil's proposed risk level classification method: A (no risk), B (negligible risk), C (minor risk), D (moderate risk), and E (major risk)[9]. As shown in Table 1.

**Table 1.** Risk level

| Risk level | description |
|---|---|
| A | No risk |
| B | negligible risk |
| C | low-level risk |
| D | moderate risk |
| E | elevated risk |

### 3.2 Construction of the Evaluation Model

Firstly, the Analytic Hierarchy Process is employed to address a intricate, multi-faceted decision-making issue as a cohesive system. The overall objective is decomposed into multiple objectives or criteria, and further into multiple indicators across several hierarchical levels. Here, we establish a three-tier hierarchical structure of Goals-Criteria-Indicators (G-C-P), as illustrated in the figure below[10]. The first tier is the top level of the entire evaluation system, representing the ultimate goal to comprehensively assess risks in cloud computing models. The second tier is the criteria attribute layer. The cabinet model decomposes the overall objective into several sub-objectives or criteria and takes data protection, access control, identity authentication, security, availability, performance, compliance, return on investment, and cost saving into consideration. The third tier is the indicator solution layer of the cabinet model. It refines the criteria attributes of the second tier by providing solution and evaluation indicators as shown in Figure 1.

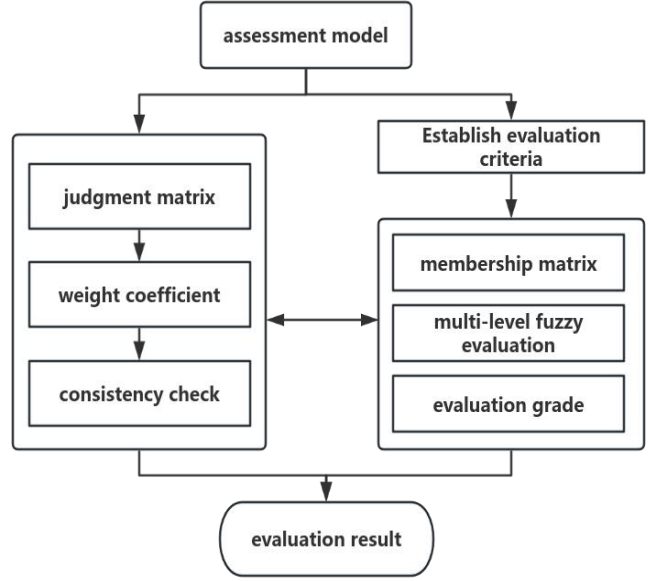

**Figure 1.** Risk assessment model

, we establish judgment matrices based on the scale from 1 to 9 to reflect the relative importance between any two risk factors. As depicted in Table2. In particular, 1 point is used for the case of equal importance between two factors, 3 points represent that one factor is slightly more important than the other, 5 points indicate that one factor is moderately more important than the other, 7 points indicate that one factor is significantly more important than the other, and 9 points indicate that one factor is extremely more important than the other. The relative importance of 2, 4, 6, and 8 between two factors lies between the relative importance of consecutive whole numbers. On this basis, we establish n×n judgment matrices with the diagonal being 1 by calculating $W_i/W_j$ for each pair of indicators, as depicted in the following formula.

$$A = \begin{bmatrix} 1 & \frac{\omega_1}{\omega_2} & \cdots & \frac{\omega_1}{\omega_n} \\ \frac{\omega_2}{\omega_1} & \cdots & \cdots & \frac{\omega_2}{\omega_n} \\ \cdots & \cdots & \cdots & \cdots \\ \frac{\omega_i}{\omega_j} & \frac{\omega_i}{\omega_j} & \cdots & 1 \end{bmatrix} = \begin{bmatrix} a_{11} & a_{12} & \cdots & a_{1n} \\ a_{21} & a_{22} & \cdots & a_{2n} \\ \cdots & \cdots & \cdots & \cdots \\ a_{n1} & a_{n2} & \cdots & a_{nn} \end{bmatrix} \quad (1)$$

$$AW = \lambda_{\max} W \quad (2)$$

**Table 2.** Satty's 9-Point Scale

| Scale | Meaning |
|---|---|
| 1 | In the comparison of the two factors, they possess equal significance.. |
| 3 | The former is slightly more important than the latter. |
| 5 | The former is significantly more important than the latter. |
| 7 | The former is strongly more important than the latter. |
| 9 | The former is extremely more important than the latter. |
| 2,4,6,8 | The middle value suggests a moderate difference in importance between the two factors. |

To verify the consistency of matrix A, we employ the Consistency Measure (CM) and the Ratio of Consistency (ROC). In this context, $\lambda_{max}$ stands for the highest eigenvalue of the evaluation matrix A. The calculated CR value should be less than 0.1 to ensure the rationality of the judgment matrix. The formulas for the calculations are as follows:

$$\lambda_{max} = \frac{(AW)_i}{n\omega_i} \tag{3}$$

$$CI = \frac{\lambda_{max} - n}{n-1} \tag{4}$$

$$CR = \frac{CI}{RI} \tag{5}$$

In cloud computing risk assessment, risk indicators are interconnected and not independent of each other. Therefore, it is necessary to comprehensively consider the influence of other indicators to avoid errors being amplified due to subjectivity and fuzziness, which can affect the results. Taking the risk factor $u_i$ as an example, we can obtain a fuzzy vector representation related to the evaluation level $v_j$:

$$R_i = (r_{i1}, r_{i2}, \cdots r_{ij}), i = 1, 2 \cdots N; j = 1, 2 \cdots n \tag{6}$$

The value range of the vector $r_{ij}$ is limited between 0 and 1, reflecting the proportional vector characteristic of the evaluation level standard $v_j$ under the risk assessment factor $u_i$. The risk assessment membership matrix R is constructed through comprehensive evaluation of n risk elements, forming an N-row and n-column matrix structure. In the membership matrix for risk assessment in a cloud computing environment, each risk factor receives an overall evaluation result. Therefore, the comprehensive evaluation of all factors in the entire risk assessment set U is reflected in the membership matrix, making it convenient to visually inspect the overall evaluation effectiveness of various audit risk factors in cloud computing under the framework of the evaluation level standard V.

$$r_1 = \begin{cases} 1 & x_i \leq v_1 \\ \dfrac{v_2 - x_i}{v_2 - v_1} & v_1 < x_i < v_2 \\ 0 & x_i > v_2 \end{cases} \tag{7}$$

$$r_2 = \begin{cases} 1 - r_1 & v_1 < x_i \leq v_2 \\ \dfrac{v_3 - x_i}{v_3 - v_1} & v_2 < x_i < v_3 \\ 0 & x_i \leq v_i 或 x_i \geq v_3 \end{cases} \tag{8}$$

$$r_j = \begin{cases} 1 - r_{j-1} & v_{j-1} < x_i \leq v_j \\ \dfrac{v_{j+1} - x_i}{v_{j+1} - v_j} & v_j < x_i < v_{j+1} \\ 0 & x_i \leq v_{j-1} 或 x_i \geq v_{j+1} \end{cases} \tag{9}$$

Under the cloud computing architecture, if we set the indicator factors for risk assessment as X, then the composition of X can be expressed as:

$$X^T = \{x_1, x_2, ..., x_m\} \tag{10}$$

Meanwhile, the classification criteria for risk assessment are denoted as V, specifically:

$$V = \{v_1, v_2, ..., v_n\} \tag{11}$$

Assuming a specific risk assessment criterion is denoted as $v_j$, and the immediate next criterion following $v_j$ is $v_{j+1}$, where $v_{j+1}$ is not less than $v_j$. For the $v_j$ level, the value $r_{ij}$ defined by its membership function is calculated based on the ith risk assessment indicator and the jth risk assessment criterion. Following this process, we can construct the cloud computing risk assessment matrix R:

$$R = \begin{bmatrix} r_{11} & r_{122} & r_{13} & \cdots & r_{1n} \\ r_{21} & r_{22} & r_{23} & \cdots & r_{2n} \\ r_{31} & r_{31} & r_{33} & \cdots & r_{3n} \\ \cdots & \cdots & \cdots & \cdots & \cdots \\ r_{m1} & r_{m2} & r_{m3} & \cdots & r_{mn} \end{bmatrix} \tag{12}$$

In the matrix, we use the letter m to represent the number of elements in the risk assessment factors, the letter n to denote the levels of risk evaluation, and $r_{ij}$ to represent the specific results of the risk assessment, where i ranges from 1 to m and j ranges from 1 to n. Assuming the total number of experts participating in the evaluation is H, then for the ith risk evaluation factor, the evaluation result of the kth expert is recorded as u(k,i,n), where k is an integer from 1 to H. In the evaluation data provided by the experts, each row contains only one 1 and the rest are 0s, and the membership matrix definition function can be expressed as 1/H. The following table shows the qualitative evaluation indicator results of a certain expert, characterized by each column containing only one 1 and the rest being 0s. The evaluation results by the experts are shown in Table 3.

$$r_{ij} = \sum_{k=1}^{H} u_{ij}^k, i = 1,2\cdots,m; j = 1,2,\cdots,n \quad (13)$$

**Table 3.** Expert Assessment Results

| object | 1 | 2 | 3 | ... | n |
|---|---|---|---|---|---|
| 1 | 0 | 1 | 0 | 0 | 0 |
| 2 | 0 | 0 | 1 | 0 | 0 |
| 3 | 1 | 0 | 0 | 0 | 0 |
| ... | 0 | 0 | 0 | 0 | 1 |
| m | 0 | 0 | 0 | 1 | 0 |

## 4. Results and discussion

The effectiveness of the proposed cloud computing security evaluation model, which integrates the Analytic Hierarchy Process (AHP) with fuzzy comprehensive evaluation, is validated in this chapter through experimental analysis. The experiment selects three typical cloud service providers (labeled as Cloud Service Provider A, B, and C respectively) as the evaluation targets and uses the indicator system constructed in this paper to assess their security capabilities. Five experts with backgrounds in cloud computing security were invited to participate in the experiment. Based on the three-tier "Objective-Criterion-Indicator" evaluation system established in Chapter 3, they scored the specific performances of the three cloud service providers across nine criteria, including data protection, access control, identity authentication, availability, and compliance. The evaluation set adopts a five-level risk rating scale (A-E), as shown in Table 4, and quantifies it into numerical values: A=5, B=4, C=3, D=2, E=1. Judgment matrices were constructed using the expert scores to calculate the weights of each criterion and indicator. Taking the "data protection" criterion as an example, the judgment matrix for its three indicators is as follows:

**Table 4.** Judgment matrix

| | Encryption capability | Backup mechanism | Data isolation |
|---|---|---|---|
| Encryption capability | 1 | 3 | 5 |
| Backup mechanism | 1/3 | 1 | 2 |
| Data isolation | 1/5 | 1/2 | 1 |

The calculated maximum eigenvalue $\lambda_{max}$ is 3.038, and the consistency ratio CR is 0.033, which is less than 0.1, indicating that it passes the consistency test. The weights of some criteria and indicators are shown in Table 5:

**Table 5.** Weight distribution of partial criteria and indicators

| Criteria | Weight | Indicator | Local weight | Global weight |
|---|---|---|---|---|
| Data protection | 0.35 | Encryption capability | 0.65 | 0.2275 |
| | | Backup mechanism | 0.23 | 0.0805 |
| | | Data isolation | 0.12 | 0.0420 |
| access control | 0.2 | Permission granularity | 0.5 | 0.1000 |
| | | audit log | 0.3 | 0.0600 |
| | | dynamic policy | 0.2 | 0.0400 |

Uitgezonden op basis van de scores van de deskundenen een membreschaperdeematreeks opgesteld en de drie Cloud-serviceleveranciers grondiger beoordeeld. Volgens de evaluatie lopen de drie Cloud-serviceleveranciers zeer voor en naach: Cloud Service Provider C behaalde een toptoesperscore en behoort daarmee met een geheel getal naar de A-klas (risk-free); Cloud Service Provider A behaalde een licht puntelager geheel getal en behoort naar de A-klas (risk-free); Cloud Service Provider B behaalde een lager geheel getal en behoort naar de B-klas (negligible risk). De toevlakking van de resultaten is in het algemeen overaccordende met de feiten die in de branche worden geobserveerd en goedenis de effectiviteit van de model-validation.Om de stabieliteit van de modeltoepassing te bepalen, veranderden we de gewichten van twee kriteria, "data protection" en "access control" met ±10%. The sensitivity analysis indicates that the relative order of the three cloud service providers does not change when the variation of the weight is within the range of fluctuation.These results indicate that the model demonstrates robustness against variations in user preferences.

## 5. Conclusion

This paper focuses on the issues related to the cloud computing security field, using the Analytic Hierarchy Process (AHP) to identify risk assessment sub-objectives and to allocate weights to the various factor indicators within the cloud computing context; subsequently, the Fuzzy Comprehensive Evaluation Technique (FCET) is employed to comprehensively evaluate the risks in the cloud computing environment. This study integrates the merits of AHP and FCEM and emphasizes the influence of multiple factors in the cloud computing risk assessment. It realizes the coordination of various kinds of analysis, reflects the fuzzy characteristics of the cloud computing risk assessment factors and process accurately, and provides a clear model framework and methodological basis to reduce the cloud computing audit risks and strengthen the audit system robustness.